\documentclass[aps,prl,twocolumn,superscriptaddress,showpacs,citeautoscript]{revtex4-1}

\usepackage{amsmath,amssymb}
\usepackage{bm}
\usepackage{graphicx}
\usepackage{epstopdf}
\usepackage{latexsym}
\usepackage{subfigure}
\usepackage{color}
\usepackage{dsfont} 
\usepackage{wasysym} 
\usepackage{tabularx}
\newcolumntype{N}{@{}m{0pt}@{}}

\newcolumntype{C}[1]{>{\centering\let\newline\\\arraybackslash\hspace{0pt}}m{#1}}

\usepackage{amsfonts}  
\usepackage{enumitem}
\usepackage{array}
\usepackage{hyperref}
\usepackage{braket}

\newcommand{\beq}{\begin{eqnarray*}}
\newcommand{\eeq}{\end{eqnarray*}}

\newcommand{\be}{\begin{eqnarray}}
\newcommand{\ee}{\end{eqnarray}}

\def\lsim{\mathrel{\rlap{\lower4pt\hbox{\hskip1pt$\sim$}}
    \raise1pt\hbox{$<$}}}                

\def\gsim{\mathrel{\rlap{\lower4pt\hbox{\hskip1pt$\sim$}}
    \raise1pt\hbox{$>$}}}                

\begin{document}

\title{Synthetic gauge fields stabilize a chiral spin liquid phase}
\author{Gang Chen}
\altaffiliation[Current address: ]{chggst@gmail.com, Department of Physics, University of Toronto, Toronto, Ontario M5S 1A7, Canada}
\affiliation{Department of Physics, University of Colorado-Boulder, Boulder, Colorado 80309-0440, USA}
\author{Kaden R.~A. Hazzard} 
\affiliation{Department of Physics, Rice University, Houston, Texas 77005, USA} 
\author{Ana Maria Rey}
\affiliation{JILA and Department of Physics, University of Colorado-Boulder, NIST, Boulder, Colorado 80309-0440, USA}
\affiliation{Center for Theory of Quantum Matter, University of Colorado, Boulder, Colorado 80309, USA}
\author{Michael Hermele}
\affiliation{Department of Physics, University of Colorado-Boulder,  Boulder, Colorado 80309-0440, USA}
\affiliation{Center for Theory of Quantum Matter, University of Colorado, Boulder, Colorado 80309, USA}

\date{\today}

\begin{abstract}
We calculate the phase diagram of the SU($N$) Hubbard model describing fermionic alkaline earth atoms in a square optical lattice with on-average one atom per site, using a slave-rotor mean-field approximation. We find that the chiral spin liquid predicted for $N\ge5$  and large interactions 
passes through a fractionalized state with a spinon Fermi surface as interactions are decreased before transitioning to a weakly interacting metal.  We also show that by adding an artificial uniform magnetic field with flux per plaquette $2\pi/N$,
the chiral spin liquid becomes the ground state for all $N\ge 3$
at large interactions, persists to weaker interactions, and its spin gap increases, suggesting that the spin liquid physics will persist to higher temperatures.  We  
discuss potential methods to realize the artificial gauge fields
and detect the predicted phases.
\end{abstract}

\maketitle

\begin{figure*}
\includegraphics[width=1.9\columnwidth,angle=0]{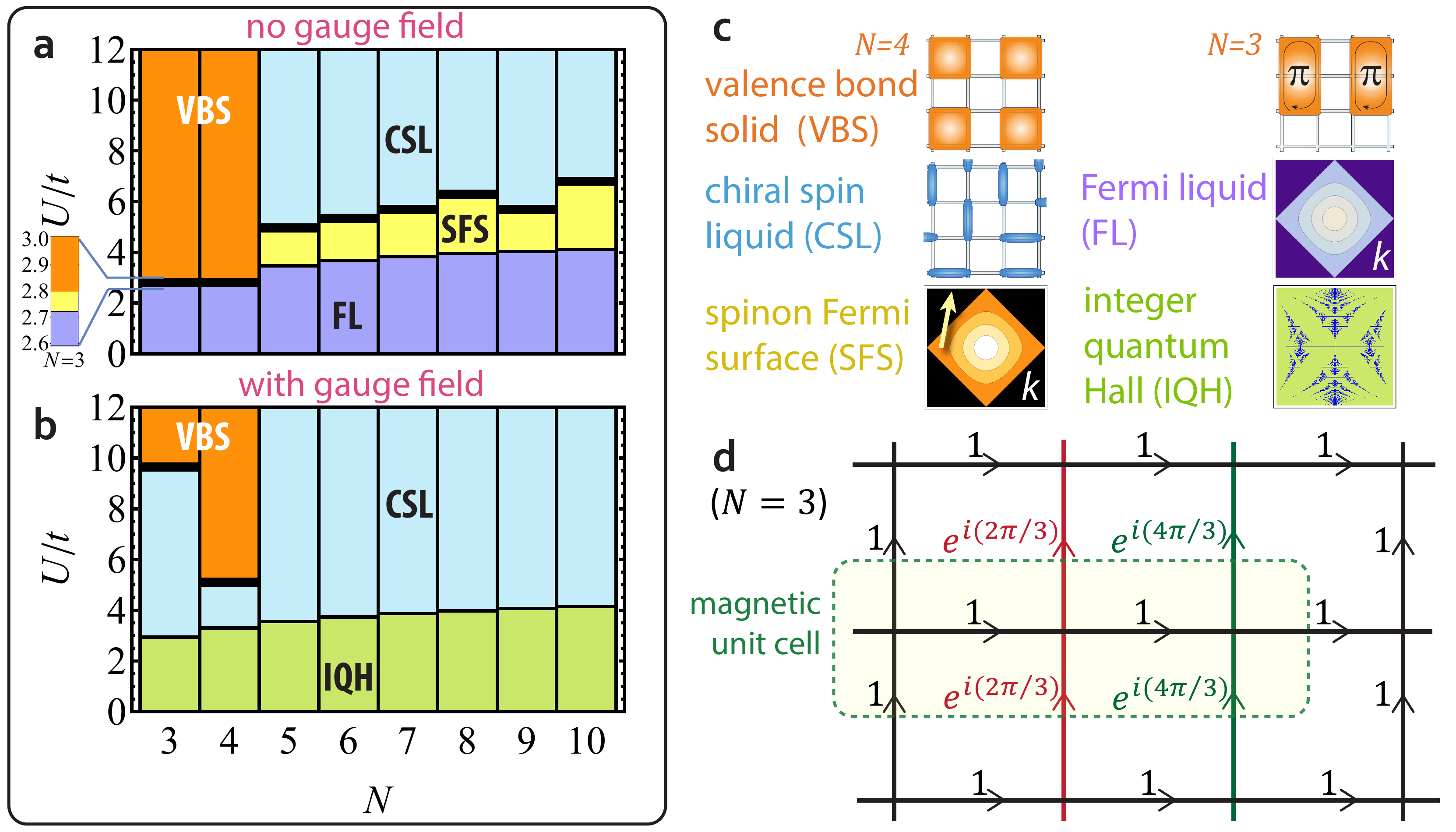}
\caption{  Phase diagram, calculated with a slave-rotor mean-field approximation, as a function of spin degrees of freedom $N$ and interaction strength $U/t$ in the (a) absence and (b) presence of an artificial uniform magnetic field with flux per plaquette $\Phi=2\pi/N$, illustrated in panel (d) for $N=3$. Thin black lines are second order phase transitions, while thick black lines are first order phase transitions. The states found are the valence bond solids (VBS), chiral spin liquid (CSL), spinon Fermi surface (SFS), Fermi liquid (FL), and integer quantum Hall (IQH) states. These are described in the text and illustrated in panel (c).    
 \label{fig:slave-rotor-phase-diagram}}
\end{figure*}

\textit{Introduction.}---The experimental realization of a topologically ordered phase of matter other than the fractional quantum Hall effect that occurs in two-dimensional electron gases is a major goal in both condensed matter and atomic physics. Phases with intrinsic topological order~\footnote{Here we mean intrinsic topological order in the sense of gapped states with long range quantum entanglements, which excludes, for example, topological band insulators.} are of fundamental interest, as they exist outside of the standard symmetry-breaking framework for classifying phases of matter and display exotic phenomena such as fractionalized excitations and edge states that are robust to local perturbations~\cite{wen:quantum_2007}; in some cases these phases have been predicted to be useful for topological quantum computation~\cite{kitaev:fault-tolerant_2003,nayak:non-abelian_2008}. 
Ultracold atomic systems are uniquely tunable and clean systems that offer a platform to realize exotic phases. 
However, so far, reaching the required low temperatures remains a challenge.  
Previous work predicted a topologically ordered chiral spin liquid (CSL) ground state in fermionic alkaline earth atoms (AEA) in a deep square optical lattice~\cite{hermele:mott_2009,hermele:topological_2012}. 
In this Letter we show, within a slave-rotor approximation,
that by applying a synthetic gauge field to this system it is possible to enhance the parameter space where the CSL exists, to increase the corresponding spin gap, and in turn to increase the temperatures at which CSL physics manifests.  
In addition, without a synthetic gauge field, away from the strongly insulating limit we find a gapless quantum spin liquid with a spinon Fermi surface.
 
Recently, experiments have trapped and cooled AEA
to quantum degeneracy and loaded them in an optical lattice \cite{takasu_spin-singlet_2003,fukuhara_bose-einstein_2007,kraft_bose-einstein_2009,stellmer_bose-einstein_2009,de_escobar_bose-einstein_2009,fukuhara_mott_2009,mickelson_bose-einstein_2010,desalvo_degenerate_2010,tey_double-degenerate_2010,taie_realization_2010,stellmer:laser_2013}. Moreover, experiments \cite{stellmer:detection_2011,pagano:one-dimensional_2014,scazza:observation_2014} have confirmed the predicted SU($N$) spin symmetry in the collisional properties of fermionic AEA \cite{wu:exact_2003,cazalilla:ultracold_2009,gorshkov:two-orbital_2010}.
This SU($N$) symmetry generalizes the usual SU($2$) symmetry, and $N$ can be controllably varied by initial state preparation up to $2I+1$, 
with $I$ the nuclear spin (as large as $N=10$ for $^{87}{\rm Sr}$ with $I=9/2$).
The low temperatures reached in recent experiments~\cite{taie:su6_2012,cazalilla:ultracold_2014,
stellmer:annual_2014}, at which short range spin correlations should begin to develop, makes it particularly timely to study quantum magnetism in these systems. Several theory works have addressed questions related to the expected SU($N$) magnetic phases in the strongly interacting limit~\cite{assaraf:sun_1999,hermele:mott_2009,toth:sun_2010,manmana:sun_2011,corboz:sun_2011,
rapp:sun_2011,nonne:sun_2011,hermele:topological_2012,bonnes:adiabatic_2012,
messio:sun_2012,
corboz:sun2_2012,corboz:sun3_2012,bauer:sun_2012,
cai:quantum_2013,wu:pomeranchuk_2013,bluemer:mott_2013,song:mott_2013,
wang:competing_2014,zhou:quantum_2014}.

In parallel, other ultracold atom experiments have realized synthetic gauge fields~\cite{lin:synthetic_2009,dalibard:artificial_2011,
aidelsburger:experimental_2011,struck:tunable_2012,
goldman:light-induced_2013,aidelsburger:realization_2013,
miyake:realizing_2013,jotzu:experimental_2014}.  
In these experiments, the atoms behave as if they were charged particles in external electromagnetic fields despite their neutrality.  Although many schemes in principle can create the gauge field that we study in this paper, we focus on methods utilizing laser-induced tunneling~\cite{Ruostekoski:particle_2002,jaksch:creation_2003,gerbier:gauge_2010}.

\textit{AEA in optical lattices with synthetic gauge fields.}---AEA in a sufficiently deep optical lattice are described by an SU($N$) generalization of the usual ($N=2$) Hubbard model, 
\be
H &=& -t\!\sum_{\langle i,j\rangle, \alpha}\! e^{ i \phi_{ij} } c^\dagger_{\alpha,i}c_{\alpha,j}^{\phantom \dagger}+\frac{U}{2}\sum_{i}   (n_{i}-1)^2 \label{eq:SUN-Hubbard}
\ee
where $c_{\alpha,i}$ is the fermionic annihilation operator for nuclear spin state $\alpha$ at lattice site $i$, $\sum_{\langle i,j\rangle}$ indicates a sum over nearest neighbors $i$ and $j$; $\phi_{ij}=-\phi_{ji}$ is the (externally imposed) lattice gauge field. We define $n_i=\sum_\alpha c^\dagger_{\alpha,i} c_{\alpha,i}^{\phantom \alpha}$, and $t$ and $U$ are the hopping energy and on-site interaction energy, whose ratio can be tuned by modifying the optical lattice depth. In this Letter, we take the average fermion number per site to be one. 

The gauge field $\phi_{ij}$ depends both on the artificial electromagnetic field as well as the gauge choice.  We are interested in the physics of a two-dimensional square lattice with a spatially uniform, time-independent artificial magnetic field, and use the Landau gauge where
\be
\phi_{ij} &=& \begin{cases} \Phi x_j \delta_{ y_j -1, y_i}    & \text{if $\{i,j\}$ bond is vertical}\\
0 & \text{otherwise},
\end{cases}
\label{eq:gauge-field}
\ee
$x_j$ is the x coordinate of site $j$ measured in lattice units, and $\Phi$
is the flux penetrating a single square plaquette of the lattice~\footnote{This is a non-dimensionalized flux: it is divided by a unit flux quantum so $\Phi$  is identical to the phase acquired around a single plaquette.}.
We focus on the case 
$\Phi=2\pi/N$, because this choice of $\Phi$ is favorable for the existence of the chiral spin liquid.
We note that the magnetic unit cell associated with the translational invariance of the Hamiltonian is enlarged from the one imposed by the optical lattice potential. Figure~\ref{fig:slave-rotor-phase-diagram}(d) shows the system with this flux and gauge choice, and the enlarged magnetic unit cell, for $N=3$.

We calculate the phase diagram and properties of this system within a slave rotor mean-field approximation \cite{Florens04,Lee05}, which we describe briefly. 
This technique is designed to match on to the previous large-$N$ solution 
in the large $U/t$ limit, and is well-suited for describing non-magnetic ground states
in proximity to the Mott transition. 
First we
expand the Hilbert space to include a U(1) bosonic rotor degree of freedom on each site, $\theta_j$, and new fermionic spinon degrees of freedom associated with operators $f_{\alpha,j}$, which are defined by 
\be 
c_{\alpha,j} &=& e^{-i\theta_j}f_{\alpha,j}.
\ee
In order to reproduce the original Hilbert space, we must impose the constraint
\be
L_j &=& \sum_\alpha f^\dagger_{\alpha,j}f^{\phantom\dagger}_{\alpha,j}-1 \label{eq:ang-mom-constraint}
\ee
that the rotor angular momentum $L_j$ is uniquely determined by the particle number. Here, $L_j$ satisfies $[\theta_j,L_j]=i$. We rewrite the Hamiltonian in terms of these new degrees of freedom, giving
\be 
H=-t\sum_{\langle i,j \rangle, \alpha} e^{i\phi_{ij}} e^{i(\theta_i-\theta_j)}f^{\dagger}_{\alpha,i}
f^{\phantom\dagger}_{\alpha,j} + \frac{U}{2} \sum_i L_i^2.
\label{eq:SUN-slave-rotor}
\ee

Although the rewritten Hamiltonian Eq.~\eqref{eq:SUN-slave-rotor} together with the constraint Eq.~\eqref{eq:ang-mom-constraint} is exactly equivalent to Eq.~\eqref{eq:SUN-Hubbard}, to make further progress we make a mean-field approximation to decouple the rotor and spinon degrees of freedom. We then obtain the coupled mean-field Hamiltonians for the rotors and 
the spinons, 
\begin{eqnarray}
H_r &=& -\sum_{\langle i,j \rangle} J_{ij} e^{i\theta_i - i\theta_j} + \sum_i \frac{U}{2} L_i^2 + h_i (L_i + 1), \\ 
H_f &=& - \sum_{\langle i,j \rangle, \alpha} \tilde{t}_{ij} e^{i\phi_{ij}} f^{\dagger}_{\alpha,i}
f^{\phantom\dagger}_{\alpha,j} - \sum_{i,\alpha} h_i f^{\dagger}_{\alpha,i}
f^{\phantom\dagger}_{\alpha,i}, 
\end{eqnarray}
where $h_i$ is a Lagrange multiplier that enforces on average the constraint Eq.~\eqref{eq:ang-mom-constraint}, 
$\tilde{t}_{ij} \equiv t \langle e^{i\theta_i - i\theta_j} \rangle_r$, and
$J_{ij} \equiv t  e^{i\phi_{ij}} \sum_{\alpha}
\langle f^{\dagger}_{\alpha,i} f^{\phantom\dagger}_{\alpha,j} \rangle_f$. 
Here the sub-index $r$ ($f$) refers to taking the expectation value in the 
rotor (spinon) mean-field ground state $|\psi\rangle_r$ ($|\psi\rangle_f$). 
The Hamiltonians $H_r$ and $H_f$ are invariant under a U(1) gauge transformation, $f^\dagger_{\alpha,i} \rightarrow f^\dagger_{\alpha,i} e^{-i \chi_i},
\theta_i \rightarrow \theta_i + \chi_i $, and $\tilde{t}_{ij} \rightarrow 
\tilde{t}_{ij} e^{i \chi_i -i \chi_j},  J_{ij} \rightarrow  J_{ij}e^{-i \chi_i + i \chi_j}$. 
We solve $H_r$ and $H_f$ self-consistently 
for several variational ansatz \cite{Supple} and find the ground state by optimizing the total energy 
$\langle \psi | H  | \psi \rangle$ where 
$H$ is given by Eq.~\eqref{eq:SUN-slave-rotor} and 
$|\psi\rangle \equiv |\psi\rangle_r | \psi\rangle_f $ 
is the mean-field state. 
 
\textit{Results.}---
Figure~\ref{fig:slave-rotor-phase-diagram}(a, b) shows the slave-rotor mean-field phase diagram as a function of $U/t$ and $N$; the top panel shows the phase diagram in the absence of a gauge field and the bottom shows the phase diagram for a gauge field with flux $\Phi=2\pi/N$. 
We find five phases: Fermi liquid (FL), integer quantum Hall (IQH), valence bond solids (VBS), a gapless spin liquid with a spinon Fermi surface (SFS) \cite{ANDERSON06031987}, and a chiral spin liquid (CSL) \cite{WenCSL1989,Laughlin1989}. Thin black lines indicate second order transitions and thick black lines indicate first order phase transitions. Generically, the role of the Hubbard $U$ interaction is to localize the atom on lattice sites. 
Such Mott localization is signalled in the rotor sector;
when the bosonic rotor is gapped and uncondensed with $\langle e^{i\theta}\rangle = 0$, the system is in a Mott insulating state. 
The mean-field parameters and some key properties of the different phases are listed in 
Table~\ref{table1}.  As we show in the table, the rotor and the spinon may experience 
different, even opposite, gauge fluxes in their mean-field Hamiltonians 
for different phases.  Since the rotor and the spinon must form a whole atom, the 
total gauge flux experienced by the rotor and the spinon  
should be equal to the synthetic gauge flux that is externally imposed on the atom. 

The FL phase is very similar to the usual SU(2) Fermi liquid, and its structure and instabilities are essentially those described in the absence of a lattice~\cite{cazalilla:ultracold_2009}. The VBS are translation-symmetry breaking phases with repeating units of SU($N$) singlets spread across multiple sites. In particular,
as we plot in Figure~\ref{fig:slave-rotor-phase-diagram}(c), the system is decoupled into 6-site rectangular (4-site square) clusters in the SU(3)-VBS [SU(4)-VBS] state. 
The SFS spin liquid state is characterized by a gapless spinon Fermi surface with a gapped bosonic rotor 
in the mean-field theory. Going beyond the mean-field description, we need to include
the U(1) phase fluctuation of the spinon hopping $\tilde{t}_{ij}$. This is
the internal gauge fluctuation \cite{Lee05};  
it is dynamically generated and 
is unrelated to the synthetic gauge field that is imposed externally.
At low energies, the SFS spin liquid is described by the spinon Fermi surface
coupled by a fluctuating internal U(1) gauge field \cite{Lee05,LeeNagaosa92,SungSik2008,SungSik2009,Mross2010,SungSik2013}. 
Due to the spinon-gauge coupling, the overdamped
 U(1) gauge fluctuation scatters the spinons on the Fermi surface
 and destroys the coherence of the spinon quasi-particles. 
The resulting state is a non-Fermi liquid of fermionic spinons. 
The CSL is distinct from the SFS in  that the spinons form an integer quantum Hall
state in the CSL. Upon coupling to U(1) gauge fluctuations, this leads to a chiral 
topologically ordered phase with anyon excitations, and gapless chiral edge states 
that carry spin but no charge \cite{WenCSL1989}.

\begin{table}
\begin{tabular}{C{1.7cm}C{1.5cm}C{1.5cm}C{1.6cm}C{1.6cm}N}
Phases & $\langle e^{i\theta} \rangle $ & rotor flux  & spinon gap & spinon flux &
\\[1.4ex] 
\hline 
FL & $\neq 0$ & 0 & 0 & 0 &
\\ [1.4ex]
SFS & 0 & 0 & 0 & 0 &
\\ [1.4ex]
CSL & 0 & $-2\pi/ N$ & $\neq 0$ & $2\pi/N$ &
\\ [1.4ex]
SU(3)-VBS & 0 & $-\pi$ & $\neq 0$ & $\pi$ &
\\[1.4ex]
SU(4)-VBS & 0 & 0 & $\neq 0$ & 0 &
\\[1.4ex]
\hline
IQH & $\neq 0$  &  0  & $\neq 0$ & $2\pi/N $ &
\\[1.4ex]
CSL & 0 & 0 & $\neq 0 $ & $2\pi/N $&
\\[1.4ex]
SU(3)-VBS & 0 & $\pi/3$ & $\neq 0$ & $\pi$ &
\\[1.4ex]
SU(4)-VBS & 0 & $\pi/2$ & $\neq 0$ & 0&
\\[1.4ex]
\hline
\end{tabular}
\caption{Parameters that characterize the obtained phases. 
The upper five (lower four) rows describe phases in the absence (presence) of the synthetic gauge field.  
The rotor (spinon) flux refers to the flux   
that is experienced by the rotor (spinon) in the mean-field Hamiltonian $H_{r}$ ($H_f$). 
For the FL, SFS, IQH, and CSL states, 
the flux is defined for the elementary square plaquette. 
For SU(3)-VBS [SU(4)-VBS] state, the flux is defined through
 the 6-site [4-site] cluster \cite{Supple}.}
 \label{table1}
\end{table}

To understand the global structure of the phase diagram, it is useful to consider the two limits $U/t=0$ and $U/t\rightarrow \infty$. The FL and IQH states are simply the non-interacting ground states occurring at $U/t=0$. In the strongly interacting limit, 
the Hubbard model reduces to an SU($N$) Heisenberg model, and the phase diagram coincides with previous slave-fermion mean-field calculations of the Heisenberg model~\cite{hermele:mott_2009}: for $N=3,4$ the ground state is a VBS, while for $N\ge 5$ the ground state is a CSL. This is true both with and without a synthetic gauge field, as in the the $U/t\rightarrow \infty$ limit the physics is governed by two-site nearest neighbor superexchange, which is insensitive to the gauge flux.

In the intermediate $U/t$ regime, the gauge field causes more significant differences. Without a gauge field, we find that an SFS phase intervenes between the non-interacting FL and Heisenberg-limit CSL or VBS for all $N$ except $N=4$, in which case there is a direct transition between the FL and VBS ground states. The FL-SFS transition is second order and is expected to remain continuous beyond mean-field theory \cite{Senthil08July1}, while the SFS-CSL and FL-VBS are first order phase transitions. In contrast, in the presence of the $\Phi=2\pi/N$ gauge flux, a direct second order transition occurs between the non-interacting IQH phase and the CSL phase
within our mean-field theory, and the CSL exists at intermediate $U/t$ even for $N=3$ and $4$.

\begin{figure}
\setlength{\unitlength}{1.0in} 
\includegraphics[width=3.in,angle=0]{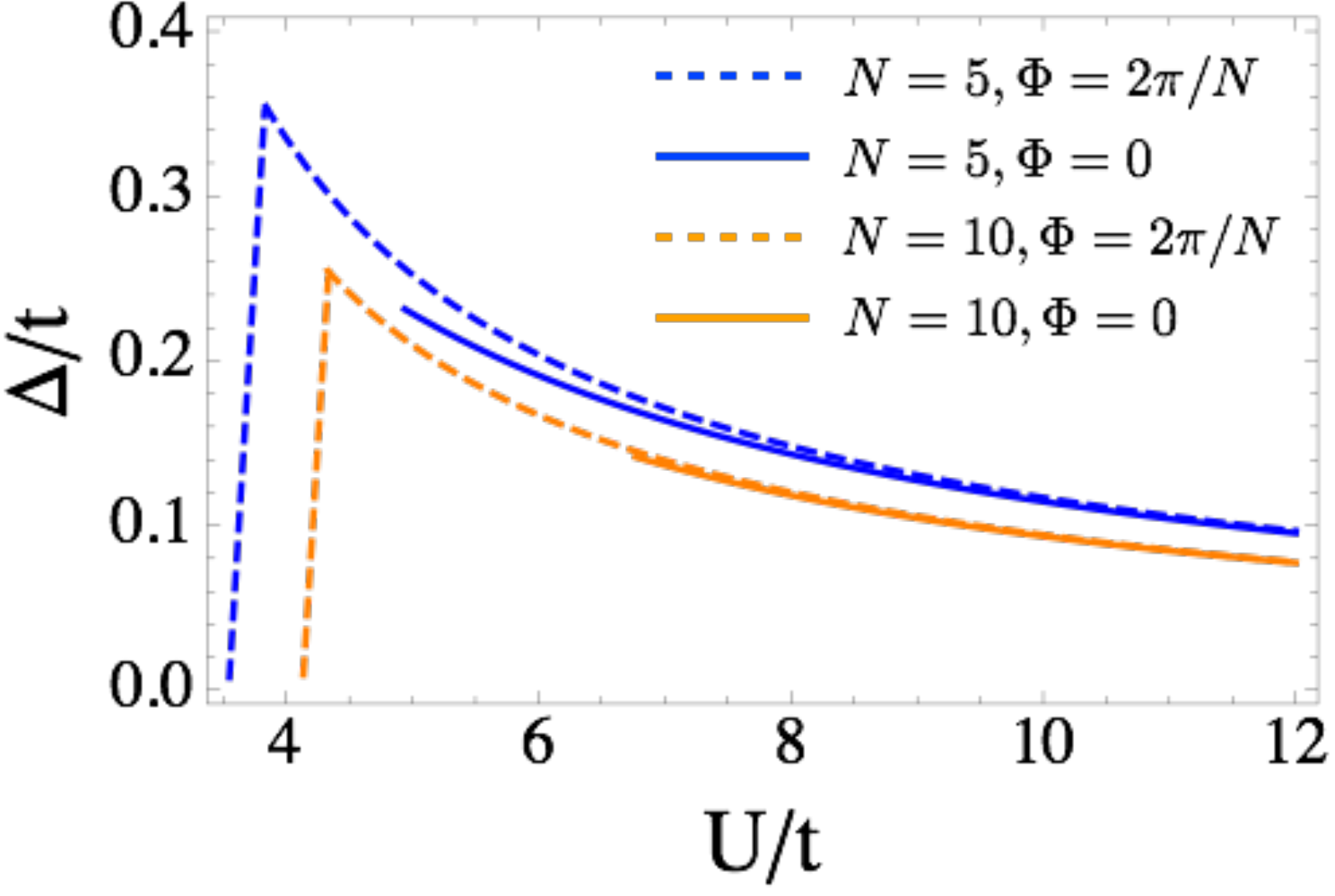}
\caption{The excitation gap of the CSL phase, $\Delta$, as a function of interaction strength, $U$, both in units of the tunnelling $t$ . 
The curves illustrate the $N$- and magnetic flux $\Phi$-dependence. From bottom to top, we show $(N=10, \Phi=0)$; $(N=10, \Phi=2\pi/10)$; $(N=5, \Phi=2\pi/5)$; and $(N=5,\Phi=2\pi/5)$.
The turning points at $U/t \approx 4$ are the locations below which the rotor gap becomes smaller than the spin gap. 
 \label{fig:spin-gaps}}
\end{figure}

The gauge field increases the parameter space for which the CSL occurs: in addition to persisting down to $N=3,4$, the CSL  occurs for a broader range of $U/t$ values. In particular, the minimum  $U/t$ for which the CSL exists decreases from about $U/t\approx5.5$ to $U/t\approx3.5$ (the exact values depend on $N$). 

In the CSL, both the spinon sector and the rotor sector are gapped. 
Figure~\ref{fig:spin-gaps} illustrates the excitation gap $\Delta$'s dependence on 
$U/t$, $N$, and the gauge flux in the CSL where $\Delta$ is the 
smaller of the spin gap and the rotor gap. 
In the slave-rotor mean-field approximation, the spin gap is 
simply the band gap of the spinon spectrum, and the rotor gap 
is set by the Hubbard $U$ interaction and thus
stays much larger than the spin gap in the Mott insulating regime
except near the Mott transition. 
For a given $U/t$, the spin gap slightly increases when the gauge field is turned on. 
An even more favorable effect of the gauge field for the spin gap occurs because the CSL persists to lower $U/t$. Since $\Delta$ increases as $U/t$ decreases, the gauge field increases the maximum $\Delta$ by about a factor of 1.5. Because $\Delta$ sets the temperature to which the CSL's characteristics remain, we therefore expect the gauge field to increase the temperature range over which the CSL behavior is accessible. 

\textit{Gauge field implementation.}---Many proposals to implement artificial gauge fields exist. Here we suggest one scheme, which uses Raman-induced tunneling in deep lattices subject to a uniform potential gradient~\cite{aidelsburger:realization_2013,
miyake:realizing_2013}. A Raman process is on resonant with the energy splitting between adjacent lattice sites, and the atoms acquire a phase kick each time they hop, imprinting the phase $\phi_{ij}$ in Eq.~\eqref{eq:SUN-Hubbard}. This scheme is 
natural for our current considerations, since it utilizes the optical lattice and generates the Hamiltonian Eq.~\eqref{eq:SUN-Hubbard} with strong gauge fluxes. 
Gauge fields have been recently demonstrated in bosonic alkali atoms using this technique~\cite{aidelsburger:realization_2013,
miyake:realizing_2013}, although we note that these experiments have observed unexplained heating, which could be problematic for realizing low temperature phases.  

We also mention the alternative scheme proposed in Ref.~\cite{gerbier:gauge_2010} that seems natural for the present work with AEA: rather than using Raman lasers, one traps the ${}^1$S$_0$ ground ($g$) and ${}^3$P$_0$ excited ($e$) states in, for example, a checkboard pattern in an optical lattice by using an appropriate, ``anti-magic," wavelength~\cite{daley:quantum_2008}. The $e$ state has a $\sim 100$s natural lifetime, and is therefore stable on the timescale of the system. Because a single laser can directly drive tunneling of a $g$ atom to an $e$ atom at an adjacent lattice site while imprinting a phase $\phi_{ij}$, one avoids the complexity of driving Raman processes. However, when this proposal is implemented in the context of interacting quantum phases
additional considerations arise that were not accounted for in the prior analysis. First, two $e$-state atoms on the same site can inelastically collide and be lost from the trap. We have found that this problem can be largely mitigated when using a checkerboard $g$-$e$ pattern \cite{Kaden}. Second, the interactions are inhomogeneous, being different for the sites occupied by $g$ atoms and $e$ atoms. 
This issue can modify the discussed phase diagram. Third, the flux generated 
in the simplest implementation of this proposal is staggered and thus requires 
rectification techniques to make it homogeneous. 

\textit{Preparation and detection.}---Reaching the temperature regimes to observe the phase diagram Figure~\ref{fig:slave-rotor-phase-diagram} is challenging.
However, the expected advantage of the SU($N$) symmetry for cooling~\cite{hazzard:high_2012,bonnes:adiabatic_2012,taie:su6_2012,wu:pomeranchuk_2013} together with the less stringent temperature requirements to observe CSL phases in the presence
of the synthetic gauge field
might help achieve the required conditions. Other potentially favorable aspects of the gauge field are the absence of an intermediate SFS phase and that all transitions are second order in the mean-field analysis. Consequently, adiabatically going from weak to strong interactions may be easier than in the absence of the gauge field.
On the other hand, the gauge field itself introduces further constraints such
as the requirement to use a deep lattice potential and a complex band structure even in the weakly interacting regime. Consequently,
determining optimal preparation is beyond the scope of this work.  

To conclude, we briefly outline methods to detect the CSL and SFS. Although it is premature to analyze protocols in detail,  as these will depend substantially on the specific experimental implementation, it is useful to describe the basic ingredients that would be required. To detect the CSL Ref.~\cite{hermele:topological_2012} suggests methods to probe two characteristic properties of topological phases: looking for topologically protected, chiral edge currents
and introducing a weak attractive optical potential that is localized to a few lattice sites, which should bind the anyonic quasiparticles. Braiding or interfering these quasiparticles can manifest their anyonic nature. 
To detect the SFS state, one can perform spin-dependent Bragg spectroscopy
to detect the 2-spinon continuum in the dynamic spin structure factor; the most basic signature of the exotic nature of this phase is the lack of order and existence of gapless excitations. More details of the state and its excitations could be revealed by considering more structure of the spectrum, similar to that considered in Refs.~\onlinecite{PhysRevB.65.165113,PhysRevB.87.045119}.

\emph{Acknowledgements.} This work was supported by the AFOSR, AFOSR-MURI, NSF JILA-PFC-1125844, NSF-PIF-1211914, NIST and ARO (AMR) and the U.S. Department of Energy, Office of Science, Basic Energy Sciences, under Award number DE-FG02-10ER46686 (G.C. and M.H.). G.C. acknowledges NSF grant no.~PHY11-25915 for supporting the visitor program at the Kavli institute for theoretical physics during the workshop ``Frustrated Magnetism and Quantum Spin Liquids'' in 2012, when and where part of the current work was done.  

\bibliography{CSL-gauge}

 \vspace{1cm}

\appendix
\begin{center}
{\bf SUPPLEMENTARY MATERIAL}
\end{center}

\section{Slave rotor mean-field theory: no translational symmetry breaking} 
\label{sec:ssec1}

\emph{Synthetic gauge field case.}---Here we give a detailed description of the slave rotor mean-field theory in the presence of the synthetic gauge flux. 
To study the energetics as well as the phase transition from the IQH to the CSL in 
Figure~\ref{fig:slave-rotor-phase-diagram}(b),  
we first choose the variational ansatz for the IQH and the CSL such that
$\tilde{t}_{ij} = \tilde{t}, J_{ij} = J $.  Moreover, we assume an uniform 
Lagrange multiplier such that $h_i  \equiv h$. This is equivalent to replacing the local 
constraint at every site with a global constraint.
This simplification is justified by the fact that 
both the IQH and the CSL preserve the lattice translation symmetry. 
The rotor and spinon mean-field Hamiltonians are then given by 
\begin{eqnarray}
H_r &=&  -J \sum_{\langle i,j \rangle}   \Phi^{\dagger}_i \Phi^{\phantom\dagger}_j  + \sum_i \frac{U}{2} L_i^2 + h (L_i + 1),
\label{eqhr}
\\
H_f &=& - \tilde{t} \sum_{\langle i,j \rangle,\alpha }  e^{i\phi_{ij}} f^{\dagger}_{\alpha,i} f^{\phantom\dagger}_{\alpha,j} - h \sum_{i,\alpha} f^{\dagger}_{\alpha,i} f^{\phantom\dagger}_{\alpha,i}, 
\end{eqnarray}
where we have replaced the rotor variable $e^{i\theta_i}$ by
a uni-modular operator $\Phi_i$ such that $| \Phi_i | \equiv 1 $. 
Since the operator $L_i = \sum_{\alpha} f^\dagger_{\alpha,i} f^{\phantom\dagger}_{\alpha,i} -1$, $h$ is then thought as a chemical potential. Because $\langle\sum_i  L_i \rangle = 0$, $h$ must vanish for the mean-field solutions. With a $2\pi/N$ flux per square plaquette and one fermion per site, the Hamiltonian $H_f$ gives a spinon band structure with $N$ bands. Only the lowest band is filled (for each species) and separated from the others by a gap. Moreover, the lowest spinon band has Chern number $C=1$ for each fermion flavor $\alpha$. As we shown in Table~\ref{table1}, whether the system is in the IQH or the CSL is determined by the behavior of the rotor sector.

To solve the rotor Hamiltonian $H_r$, we implement a coherent state path integral formalism in 
imaginary time. We integrate out the conjugate variable $L_i$ and obtain the partition function that is written as a functional integration over the $\Phi$ variable,
\begin{equation}
{\mathcal Z} \simeq \int {\mathcal D} \Phi^{\dagger} {\mathcal D} \Phi {\mathcal D} \lambda
e^{- S - \int_\tau \lambda_i (| \Phi_i |^2 -1 )}.
\end{equation}
Here $\lambda_i$ is the Lagrange multiplier introduced to the unimodular condition for the 
rotor variable at every lattice site. 
The effective action is given by
\begin{equation}
S = \int_{d \tau} \frac{1}{2U} \sum_{{\bf k} \in \text{BZ} }
|\partial_{\tau } \Phi_{\bf k}|^2 
 - 2J \sum_{{\bf k} \in \text{BZ} }(
\cos k_x + \cos k_y) |\Phi_{\bf k}|^2, 
\end{equation}
where ``BZ'' refers to the Brioullin zone of the square optical lattice and we have set the lattice
constant to unity. In a standard spherical approximation for a mean-field (or saddle point) analysis, we assume a uniform Lagrange multiplier $\lambda_i \equiv \lambda $. 
We integrate out the variable $\Phi$ and obtain the saddle point equation for $\lambda$,
\begin{equation}
\frac{1}{N_s}\sum_{{\bf k}\in \text{BZ}} \frac{U}{\omega_{\bf k}} = 1,
\label{eqsadd}
\end{equation}
where $\omega_{\bf k} = [2U(\lambda - 2J (\cos k_x + \cos k_y) )]^{1/2}$ is the band dispersion of the rotor
and $N_s$ is the number of lattice sites. 
We solve the saddle point equation Eq.~\eqref{eqsadd} self-consistently with the spinon mean-field Hamiltonian $H_f$. When the rotor band touches zero energy, the rotor is condensed
and the internal U(1) gauge field picks up a mass due to the Anderson-Higgs mechanism. The rotor and 
the spinon are then bound together and form a fermionic atom. The resulting phase is the IQH. 
When the rotor band is gapped and the rotor is not condensed, 
the internal U(1) gauge field is gapped out by the Chern-Simons term 
and the resulting phase is the CSL. 

In the IQH, the system has $N$ chiral edge modes that transport spin quantum numbers as well as atoms. The CSL, however, is a Mott insulating state. The atoms are localized by the interaction in the CSL. The chiral edge states in the CSL only carry spin quantum number and cannot transport charge. The effect of the phase transition from the IQH to the CSL on the edge states is to gap out the mode that transports atoms.

\begin{figure}[htp]
{\includegraphics[width=3.5cm]{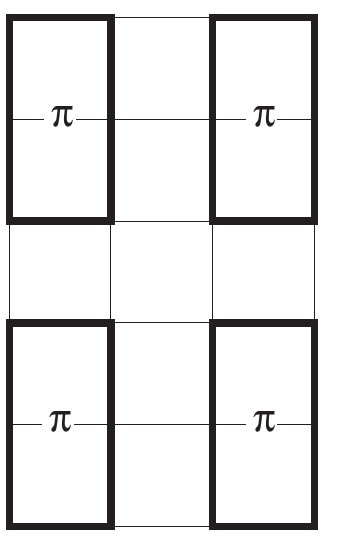}}
\caption{ The VBS state for the SU(3) model. The spinon hopping 
is zero on the light bonds and is non-zero on the dark bond. 
The $\pi$ flux through the bold rectangle is the mean-field gauge flux 
felt by the fermionic spinons in the mean-field Hamiltonian $H_f$. The figure is adapted from Ref.~\onlinecite{hermele:topological_2012}.}
\label{sfig1}
\end{figure}

\emph{No synthetic gauge field case.}---With no gauge flux the Hamiltonian is modified by putting $\phi_{ij} \rightarrow 0$.
For the FL and the SFS, we choose the spinon mean-field ansatz such that $\tilde{t}_{ij}   \equiv \tilde{t}, J_{ij} \equiv J$ and $h_i \equiv h$. 
The spinons partially fill the bands and give rise to the spinon Fermi surface. Again, whether the system is in the FL or the SFS is determined 
by the rotor sector. 
Since the FL and the SFS only occur for the model without the synthetic gauge flux, 
the rotor sector Hamiltonian is identical to Eq.~\eqref{eqhr}. 
When the rotor is condensed, the system falls into the FL. When the rotor is gapped, the system is in the SFS whose low energy property is described by the spinon Fermi surface coupled with a fluctuating U(1) gauge field.

\section{Variational ansatz: VBS states}
\label{sec:ssec2}

\begin{figure}[ht]
{\includegraphics[width=3.5cm]{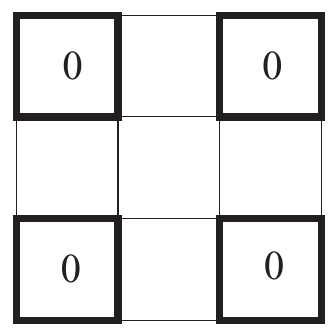}}
\caption{ The VBS state for the SU(4) model. The spinon hopping 
is zero on the light bonds and is non-zero on the dark bond. 
The $0$ flux through the bold square is the mean-field gauge flux 
felt by the fermionic spinons in the mean-field Hamiltonian $H_f$. The figure is adapted from Ref.~\onlinecite{hermele:topological_2012}. }
\label{sfig2}
\end{figure}

As we described in the main text, VBS states
become favorable in the
strongly interacting limit
 for the SU(3) and SU(4) models.
As expected, a similar conclusion was found in the previous slave-fermion study,
{\it i.e.} the ground state of the SU(3) [SU(4)] Heisenberg model -- the $U/t\rightarrow \infty$ limit of the Hubbard model -- on the square lattice \cite{hermele:topological_2012} favored the VBS state shown in Figure~\ref{sfig1} [Figure~\ref{sfig2}]. In our slave rotor mean-field calculation, 
we have chosen $H_f$ such that the spinons have the same hopping and feel the same mean-field gauge fluxes as those ones shown in Figure~\ref{sfig1} and Figure~\ref{sfig2}, while the rotor sector has the same cluster structure as the spinon sector but experiences a different flux (see Table.~\ref{table1}). 

\end{document}